\documentclass[prd,twocolumn,twoside,preprintnumbers,superscriptaddress,nofootinbib]{revtex4}

\usepackage{amsmath,slashed}
\usepackage{graphicx,graphics,color}
\usepackage{dcolumn}
\usepackage[hyperfootnotes=false]{hyperref}
\usepackage{xspace}
\usepackage{enumerate}
\usepackage{varwidth}

\usepackage{physics}
\usepackage{amssymb}
\usepackage{mathtools}
\usepackage{dsfont}

\newcommand{\Fpi}{F_\pi}
\newcommand{\mpi}{M_{\pi}}
\newcommand{\Order}{\mathcal{O}}
\newcommand{\MeV}{\,\text{MeV}}
\newcommand{\GeV}{\,\text{GeV}}

\newcommand{\Br}{\text{Br}}

\newcommand{\beq}{\begin{equation}}
\newcommand{\eeq}{\end{equation}}
\newcommand{\Op}{\mathcal{O}}
\newcommand{\F}{\mathcal{F}}

\newcommand{\eps}{\varepsilon}
\newcommand{\etap}{{\eta'}}

\newcommand{\Id}{\mathds{1}}

\renewcommand{\qq}{\mathbf{q}}
\newcommand{\spin}{\mathbf{S}}

\allowdisplaybreaks[1]

\begin{document}

\title{Improved limits on lepton-flavor-violating decays of light pseudoscalars\\[2mm] via spin-dependent $\boldsymbol{\mu\to e}$ conversion in nuclei}

\author{Martin Hoferichter}
\affiliation{Albert Einstein Center for Fundamental Physics, Institute for Theoretical Physics, University of Bern, Sidlerstrasse 5, 3012 Bern, Switzerland}
\author{Javier Men\'endez}
\affiliation{Department of Quantum Physics and Astrophysics and Institute of Cosmos Sciences, University of Barcelona, 08028 Barcelona, Spain}
\author{Frederic No\"el}
\affiliation{Albert Einstein Center for Fundamental Physics, Institute for Theoretical Physics, University of Bern, Sidlerstrasse 5, 3012 Bern, Switzerland}

\begin{abstract}
  Lepton-flavor-violating decays of light pseudoscalars, $P=\pi^0,\eta,\eta'\to\mu e$, are stringently suppressed in the Standard Model up to tiny contributions from neutrino oscillations, so that their observation would be a clear indication for physics beyond the Standard Model. However, in effective field theory such decays 
  proceed via axial-vector, pseudoscalar, or gluonic operators, which are, at the same time, probed in spin-dependent $\mu\to e$ conversion in nuclei. We derive master formulae that connect both processes in a model-independent way in terms of Wilson coefficients, and study the implications of current $\mu\to e$ limits in titanium for the $P\to\mu e$ decays. We find that these indirect limits surpass direct ones by many orders of magnitude.         
\end{abstract}

\maketitle

\section{Introduction}

In the Standard Model (SM) of particle physics, the flavor of charged leptons is conserved apart from tiny corrections due to nonvanishing neutrino masses. Nonetheless, neutrino oscillations contribute to charged lepton-flavor-violating (LFV) decays suppressed by the ratio of the neutrino to the $W$-boson masses $(m_\nu/M_W)^4$, with resulting  
branching ratios of order $10^{-50}$. Thus any observation of LFV in the charged sector would constitute a discovery of physics beyond the SM (BSM)~\cite{Petcov:1976ff,Marciano:1977wx,Marciano:1977cj,Lee:1977qz,Lee:1977tib}, see, e.g., Refs.~\cite{Kuno:1999jp,Mihara:2013zna,Calibbi:2017uvl} for reviews. 

The leading limits on such LFV decays are obtained from $\mu\to e$ transitions, with $\Br[\mu\to e\gamma]<4.2\times 10^{-13}$~\cite{MEG:2016leq}, $\Br[\mu\to 3e]<1.0\times 10^{-12}$~\cite{SINDRUM:1987nra} for purely leptonic processes (all limits given at $90\%$ confidence level), and 
\begin{align}
 \Br[\mu\to e, \text{Ti}]&<6.1\times 10^{-13}&&\text{\cite{Wintz:1998rp}},\notag\\
 \Br[\mu\to e, \text{Au}]&<7\times 10^{-13}&&\text{\cite{SINDRUMII:2006dvw}},
 \label{muelimits}
\end{align}
for $\mu\to e$ conversion in the field of an atomic nucleus, with branching fractions normalized to the respective rate for nuclear capture~\cite{Suzuki:1987jf}.\footnote{Reference~\cite{Wintz:1998rp} represents the final result by the SINDRUM-II experiment for $\mu\to e$ conversion in Ti, superseding the earlier limit $\Br[\mu\to e, \text{Ti}]<4.3\times 10^{-12}$~\cite{SINDRUMII:1993gxf}. We thank Peter Wintz for clarification on this point.} While leptonic limits will improve at the MEG II~\cite{MEGII:2018kmf} and Mu3e~\cite{Mu3e:2020gyw} experiments (and potentially beyond~\cite{Aiba:2021bxe}), especially significant improvements up to four orders of magnitude beyond the present limits~\eqref{muelimits} are projected for $\mu\to e$ conversion at Mu2e~\cite{Mu2e:2014fns} and COMET~\cite{COMET:2018auw}.

Independent constraints on $\mu\to e$ transitions can be obtained from LFV decays of light pseudoscalars, ${P=\pi^0,\eta,\eta'\to\mu e}$, for which the current limits read~\cite{ParticleDataGroup:2020ssz}
\begin{align}
 \Br[\pi^0\to \mu^+ e^-] &< 3.8\times 10^{-10}&& \text{\cite{Appel:2000wg}},\notag\\
 \Br[\pi^0\to \mu^- e^+] &< 3.2\times 10^{-10}&& \text{\cite{NA62:2021zxl}},\notag\\
  \Br[\pi^0\to \mu^+ e^-+\mu^- e^+] &< 3.6\times 10^{-10}&& \text{\cite{KTeV:2007cvy}},\notag\\
   \Br[\eta\to \mu^+ e^-+\mu^- e^+] &< 6\times 10^{-6}&& \text{\cite{White:1995jc}},\notag\\
   \Br[\eta'\to \mu^+ e^-+\mu^- e^+] &< 4.7\times 10^{-4}&& \text{\cite{CLEO:1999nsy}}. 
   \label{pseudoscalar_limits}
\end{align}
In particular, Ref.~\cite{NA62:2021zxl} improves the previous limit~\cite{Appel:2000tc} on the $\pi^0\to\mu^- e^+$ channel by an order of magnitude, leading to three independent constraints on $\pi^0\to \mu e$ all at the level of $10^{-10}$~\cite{Appel:2000wg,NA62:2021zxl,KTeV:2007cvy}. Limits on the analogous $\eta$, $\eta'$ decays are much weaker, but could be improved substantially at the planned JEF~\cite{JEF-PAC42} and REDTOP~\cite{REDTOP:2022slw} experiments. 

In this Letter we study the relation between LFV pseudoscalar decays~\eqref{pseudoscalar_limits} and $\mu\to e$ conversion limits~\eqref{muelimits}. In an effective-field-theory (EFT) approach to LFV~\cite{Petrov:2013vka,Crivellin:2013hpa,Hazard:2016fnc,Crivellin:2017rmk,Cirigliano:2017azj,Davidson:2018kud,Davidson:2020hkf,Davidson:2022nnl}, only axial-vector, pseudoscalar, or gluonic operators can contribute to the pseudoscalar decays, with scalar and vector operators forbidden by parity. 
Accordingly, the responses for the relevant operators only give rise to so-called spin-dependent (SD) $\mu\to e$ conversion~\cite{Cirigliano:2017azj,Davidson:2017nrp,Gan:2020aco} which is not enhanced by the coherent sum over the entire nucleus---these operators probe the spins of the nucleons, which combine in spin-zero pairs due to the nuclear pairing interaction. In addition to this lack of coherence, weaker limits are expected compared to vector or scalar operators because SD responses vanish for nuclei with even number of protons and neutrons, which are spinless. Thus only nuclei with odd number of nucleons contribute at all. Moreover, controlling the nuclear structure for a nucleus as heavy as $^{197}$Au is challenging, leaving in practice $^{47}$Ti and $^{49}$Ti, with low natural abundances of $7.44\%$ and $5.41\%$, respectively, that further dilute the interpretation of the experimental limit~\eqref{muelimits}. For these reasons, one might expect that limits derived from pseudoscalar decays could be competitive for these operators. 

To address this question systematically, we derive master formulae that express the $P\to \mu e$ branching ratio and the $\mu\to e$ conversion rate in terms of the same effective Wilson coefficients, and provide all hadronic matrix elements and nuclear structure factors required for a model-independent comparison. Since $\mu\to e$ conversion and pseudoscalar decays probe different linear combinations of Wilson coefficients, we study which $P\to \mu e$ regions in parameter space are least subject to independent limits, and comment on the role of renormalization group (RG) corrections in closing the resulting flat directions.

\section{Formalism}

The relevant operators up to dimension $7$ that can generate SD responses in $\mu\to e$ conversion are
\begin{align}
	\mathcal{L}_\text{eff}^\text{SD} &= 
\frac{1}{\Lambda^2} \sum_{\substack{Y=L,R\\q=u,d,s}} \Big[C^{P,q}_Y \qty(\overline{e_Y} \mu) \qty(\bar{q} \gamma_5 q) \notag\\
&+ C^{A,q}_Y \qty(\overline{e_Y} \gamma^\mu \mu) \qty(\bar{q} \gamma_\mu \gamma_5 q)
+C_Y^{T,q}\qty(\overline{e_Y} \sigma^{\mu\nu} \mu) \qty(\bar{q} \sigma_{\mu\nu} q)
\Big]\notag\\
&+ \frac{i\alpha_s}{\Lambda^3}\sum_{Y=L,R}\qty(\overline{e_Y} \mu)C^{G\tilde{G}}_Y G^a_{\mu\nu} \tilde{G}_a^{\mu\nu}  
	+ \text{h.c.}, 
\label{eq:QuarkLevelLagrangian}
\end{align}
while the leading spin-independent (SI) contributions arise from the analogous scalar, vector, and gluon operators (with Wilson coefficients denoted by $C^{S,q}_Y$,  $C^{V,q}_Y$, and $C^{GG}_Y$ in the following).\footnote{In the SI case there is also a contribution from the dipole operator, which we do not need for the present analysis and thus omit for simplicity.} The projectors are introduced as $\overline{e_Y}=\overline{e}P_{\bar{Y}}$, with $Y \in \{L,R\}$ and $P_{L/R} = (\Id \mp \gamma_5)/2$, to make explicit that the left- and right-handed components $e_L$ and $e_R$ decouple in the limit $m_e\to 0$, which we assume throughout this Letter. The BSM scale $\Lambda$ is introduced to make the Wilson coefficients dimensionless. 

In these conventions, the decay rate becomes
\beq
\Br[P\to \mu^\mp e^\pm] = \frac{(M_P^2 - m_\mu^2)^2}{16 \pi \Gamma_P M_P^3} \sum_{Y=L,R} \abs{C^P_Y}^2,
\label{master_Pmue}
\eeq
where the Wilson coefficients and hadronic matrix elements are combined in
\begin{align}
    C^P_Y &= \sum_q \frac{b_q}{\Lambda^2}\qty(\pm C^{A,q}_Y f^q_P m_\mu - C^{P,q}_Y \frac{h^q_P}{2 m_q})
    + \frac{4\pi}{\Lambda^3} C^{G\tilde{G}}_Y a_P,
\end{align}
and the upper/lower sign applies to $\mu^\mp e^\pm$.
The matrix elements $f_P^q$, $h_P^q$, $a_P$ are defined by~\cite{Beneke:2002jn}
\begin{align}
 \langle 0|\bar q \gamma^\mu\gamma_5 q|P(k)\rangle &= i b_q f_P^q k^\mu,\notag\\ 
 \langle 0|m_q\bar q i\gamma_5 q|P(k)\rangle &= \frac{b_qh_P^q}{2},\notag\\
 \langle 0|\frac{\alpha_s}{4\pi} G^a_{\mu\nu} \tilde{G}_a^{\mu\nu}|P(k)\rangle&=a_P,
\end{align}
with dual field strength tensor $\tilde G_a^{\mu\nu}=\frac{1}{2}\eps^{\mu\nu\alpha\beta}G^a_{\alpha\beta}$, $\eps^{0123}=+1$, and satisfy the Ward identity 
\beq
b_q f_P^q M_P^2=b_q h_P^q-a_P,
\label{Ward}
\eeq
while the matrix element of the tensor current vanishes.  
The numerical coefficients are
$b_u=b_d=1/\sqrt{2}$, $b_s=1$, and $M_P$, $m_\mu$, and $m_q$ denote the pseudoscalar, muon, and quark masses, respectively. Phenomenologically, the $2\times 3\times 3+3=21$ parameters can be further reduced using isospin symmetry and neglecting strangeness and gluonic contributions to the pion matrix elements. This leaves as free parameters the pion decay constant $F_\pi$, the singlet and octet decay constants $F^0$, $F^8$, the corresponding mixing angles $\theta_0$, $\theta_8$, as well as gluon parameters $a_0$, $\theta_y$.  The explicit parameterization reads
\begin{align}
	f^u_\pi &=-f^d_\pi= \sqrt{2}\, F_\pi,\qquad f^s_\pi=0,\qquad a_\pi=0,\notag\\
 	f^u_\eta &=f^d_\eta= \sqrt{\frac{2}{3}}F^8\cos\theta_8 - \frac{2}{\sqrt{3}} F^0\sin\theta_0,  \notag\\
 	f^s_\eta &= -\frac{2}{\sqrt{3}} F^8\cos\theta_8 - \sqrt{\frac{2}{3}} F^0\sin\theta_0,\notag  \\
 	f^u_\etap &=f^d_\etap= \sqrt{\frac{2}{3}}  F^8\sin\theta_8 + \frac{2}{\sqrt{3}} F^0\cos\theta_0,\notag  \\
	f^s_\etap &= -\frac{2}{\sqrt{3}} F^8\sin\theta_8 + \sqrt{\frac{2}{3}}  F^0\cos\theta_0,\notag\\
	a_\eta&=-a_0\sin\theta_y,\qquad a_\etap=a_0\cos\theta_y, 
\end{align}
which determines the pseudoscalar matrix elements $h_P^q$ via Eq.~\eqref{Ward}. Table~\ref{tab:matrix_elements_P} collects selected numerical values for these parameters.  

\begin{table}[t]
	\renewcommand{\arraystretch}{1.3}
	\centering
	\begin{tabular}{l r r r r r}
		\toprule
		 & $\pi$ & \multicolumn{2}{c}{$\eta$} & \multicolumn{2}{c}{$\etap$}\\
		 & & Ref.~\cite{Escribano:2015yup} & Ref.~\cite{Bali:2021qem} & Ref.~\cite{Escribano:2015yup} & Ref.~\cite{Bali:2021qem}\\\colrule
		 $\frac{b_u f_P^u}{F_\pi}$ & $1$ & $0.80$  & $0.77$ & $0.66$ & $0.56$\\ 
		 $\frac{b_d f_P^d}{F_\pi}$ & $-1$ & $0.80$ & $0.77$ & $0.66$  & $0.56$\\
		 $\frac{b_s f_P^s}{F_\pi}$ & $0$ & $-1.26$ & $-1.17$ & $1.45$ & $1.50$\\
		 $a_P\,[\text{GeV}^3]$ & $0$ & -- & $-0.017$ & -- & $-0.038$\\
		 \colrule
		 $a_P^\text{FKS}\,[\text{GeV}^3]$ & $0$ & $-0.022$ & $-0.021$ & $-0.056$ & $-0.048$\\
		\botrule
	\end{tabular}
	\caption{Numerical values for the axial-vector and gluonic matrix elements contributing to the $P\to \mu e$ decays, from a phenomenological extraction via $\eta$, $\etap$ transition form factors~\cite{Escribano:2015yup} and the recent lattice-QCD calculation~\cite{Bali:2021qem} ($\overline{\text{MS}}$ scale $\mu=2\GeV$).
	The last line indicates the value of $a_P$ extracted from $f_P^u$ in the 
	Feldmann--Kroll--Stech (FKS) scheme~\cite{Feldmann:1998vh}. 
	We use $\Fpi=92.28\MeV$~\cite{ParticleDataGroup:2020ssz}.}  
	\label{tab:matrix_elements_P}
\end{table}

A decomposition analogous to Eq.~\eqref{master_Pmue} applies to the rate for $\mu\to e$ conversion in nuclei, see Ref.~\cite{Noel} for the general form. In addition to nucleon matrix elements, these processes involving atomic nuclei depend on nuclear structure factors, which encode the structure of the many-body nuclear state. These are often included in terms of a multipole decomposition~\cite{Serot:1978vj,Donnelly:1978tz,Donnelly:1979ezn,Walecka:1995mi,Glick-Magid:2022erc}, and two-body corrections can be addressed in chiral EFT, see Refs.~\cite{Klos:2013rwa,Hoferichter:2015ipa,Hoferichter:2016nvd,Hoferichter:2017olk,XENON:2018clg,Hoferichter:2018acd,Hoferichter:2020osn}.
As a final step, the $\mu\to e$ conversion rate involves atomic wave functions, describing the bound-state physics of the initial muon in the $1S$ state of the atom as well as the overlap with the final-state electron. For the SI process, these effects have traditionally been parameterized in terms of overlap integrals~\cite{Kitano:2002mt}, where effectively only the leading $M$ multipole is kept, convolved with the solution of the Dirac equation for the electromagnetic potential of the nuclear charge distribution~\cite{DeVries:1987atn}. Keeping only scalar and vector operators, the SI branching fraction becomes
\beq
\Br_\text{SI}[\mu\to e]=\frac{4m_\mu^5}{\Gamma_\text{cap}}\sum_{Y=L,R}\Bigg|\sum_{\substack{N=p,n\\\Op=S,V}}\bar C_Y^{\Op,N} \Op^{(N)}\Bigg|^2,
\label{BR_SI}
\eeq 
where $\Gamma_\text{cap}$ is the capture rate, $\Op^{(N)}$ are the (dimensionless) overlap integrals~\cite{Kitano:2002mt}, and 
\begin{align}
    \bar{C}^{S,N}_Y &= \frac{1}{\Lambda^2}\sum_q C^{S,q}_{Y} \frac{m_N}{m_q} f^{N}_q + \frac{4 \pi}{\Lambda^3} C^{GG}_Y a_N,\notag\\
    \bar{C}^{V,N}_Y &= \frac{1}{\Lambda^2}\sum_q C^{V,q}_{Y} f^N_{V_q}, 
    \label{coefficients_SI}
\end{align}
subsume Wilson coefficients and nucleon matrix elements. At leading order in the momentum expansion only the scalar/vector couplings $f_q^N$/$f_{V_q}^N$ enter~\cite{Cirigliano:2009bz,Crivellin:2013ipa,Crivellin:2014cta,Hoferichter:2015dsa,RuizdeElvira:2017stg,Gupta:2021ahb}, while the gluon operator can be expressed via the trace anomaly of the energy-momentum tensor~\cite{Shifman:1978zn}. As the SI contribution only affects the pseudoscalar decays indirectly, via RG and relativistic corrections, it suffices to consider the leading contributions~\eqref{coefficients_SI} in this work, see Refs.~\cite{Cirigliano:2022ekw,Hoferichter:2012wf,Noel,Rule:2021oxe} for two-body and momentum-dependent corrections as well as other nuclear multipoles.  

\begin{table}[t]
	\renewcommand{\arraystretch}{1.3}
	\centering
	\begin{tabular}{l c c c c}
		\toprule
		 & $S^{(p)}$ & $V^{(p)}$ & $S^{(n)}$ & $V^{(n)}$\\\colrule
		 Ref.~\cite{Kitano:2002mt}, method 1 & $0.0368$ & $0.0396$ & $0.0435$ & $0.0468$\\
		  Ref.~\cite{Kitano:2002mt}, method 3 & $0.0371$ & $0.0399$ & $0.0462$ & $0.0495$\\
		 This work & \multicolumn{2}{c}{$0.039$} & \multicolumn{2}{c}{$0.044$}\\
		\botrule
	\end{tabular}
	\caption{Overlap integrals for $^{48}$Ti compared to Ref.~\cite{Kitano:2002mt}. We find $Z_\text{eff}=17.65$, using the charge distribution from Ref.~\cite{DeVries:1987atn} in the solution of the Dirac equation (for $^{27}$Al we have $Z_\text{eff}=11.64$). Methods 1 and 3 differ mainly in the estimate of the neutron distribution.}  
	\label{tab:overlap_integrals}
\end{table}

Approximating the electron and muon wave function by a plane wave and its average value in the nucleus, respectively, the overlap integrals become
\beq
S^{(N)}=V^{(N)}=\frac{(\alpha Z)^{3/2}}{4\pi}\bigg(\frac{Z_\text{eff}}{Z}\bigg)^2\F^{M}_N(m_\mu^2),
\label{S_D_approximation}
\eeq
where $\F^{M}_N(m_\mu^2)$ denote the structure factors for the $M$ multipole, evaluated at momentum transfer $\qq^2=m_\mu^2$ and normalized to the number of protons ($Z$) or neutrons ($N$), $\F^M_p(0)=Z$, $\F^M_n(0)=N$, and $Z_\text{eff}$ parameterizes the wave-function average~\cite{Kitano:2002mt}. For the numerical analysis we use nuclear structure factors obtained using the nuclear shell model~\cite{Caurier:2004gf,Otsuka:2018bqq} with the code ANTOINE~\cite{Caurier:1999,Caurier:2004gf}. Our calculations for Ti isotopes use the KB3G interaction~\cite{Poves:2000nw} in a configuration space consisting of the $0f_{7/2}$, $1p_{3/2}$, $1p_{1/2}$ and $0f_{5/2}$ proton and neutron orbitals, with a $^{40}$Ca core. For $^{27}$Al we use the USDB interaction~\cite{Brown:2006gx} and the $0d_{5/2}$, $0d_{3/2}$, $1s_{1/2}$ configuration space with an $^{16}$O core, see App.~\ref{app:nuclear} for details~\cite{App}. \nocite{ensdf,Richter2008,Kumar:2015zuv,Matsubara:2015tua,Martinez-Pinedo:1996zvt,Angeli:2013epw} 
In particular, for $^{48}$Ti Table~\ref{tab:overlap_integrals} compares the approximation~\eqref{S_D_approximation} to Ref.~\cite{Kitano:2002mt}, showing reasonable agreement. 
Note that differences at this level are even expected, as we rely on the neutron distribution predicted by the nuclear shell model, not the assumptions from Ref.~\cite{Kitano:2002mt}. For this work the approximation~\eqref{S_D_approximation} thus proves sufficient, and we refer to Ref.~\cite{Noel} for the full analysis.   

Under the same assumptions, the decay rate for SD $\mu\to e$ conversion can be written as  
\begin{align}
	\Br_\text{SD}[\mu \to e] &= \frac{4m_\mu^5 \alpha^3 Z^3}{\pi \Gamma_\text{cap} (2J+1)}\bigg(\frac{Z_\text{eff}}{Z}\bigg)^4\\
	&\times 
	\sum_{\substack{Y=L,R\\\tau=\mathcal{L},\mathcal{T}}}\Big[C_Y^{\tau,00}S_{00}^\tau
	+C_Y^{\tau,11}S_{11}^\tau
	+C_Y^{\tau,01}S_{01}^\tau\Big],\notag
	\label{BR_SD}
\end{align}
where $J$ is the spin of the nucleus, $S_{ij}^\tau$ are the transverse ($\mathcal{T}$) and longitudinal ($\mathcal{L}$) structure factors~\cite{Hoferichter:2020osn} (corresponding to the multipoles $\Sigma'$ and $\Sigma''$, respectively),
and the coefficients receive contributions from all operators in Eq.~\eqref{eq:QuarkLevelLagrangian}. 
Defining 
\begin{align}
    \bar{C}^{P,N}_Y &= \frac{1}{\Lambda^2}\sum_q C^{P,q}_{Y} \frac{m_N}{m_q} g^{q,N}_5 - \frac{4 \pi}{\Lambda^3}  C^{G\tilde{G}}_Y \tilde{a}_N,\\
    \bar{C}^{A,N}_Y &=\frac{1}{\Lambda^2} \sum_q C^{A,q}_{Y} g^{q,N}_A,\quad 
    \bar{C}^{T,N}_Y =\frac{1}{\Lambda^2} \sum_q C^{T,q}_{Y} f^{q,N}_{1,T},\notag
\end{align}
with nucleon matrix elements at vanishing momentum transfer in the conventions of Ref.~\cite{Hoferichter:2020osn}
\begin{align}
\langle N|\bar q\gamma^\mu\gamma_5 q|N\rangle&=g_A^{q,N}\langle N|\bar N\gamma^\mu\gamma_5 N|N\rangle,\notag\\
m_q\langle N|\bar q i\gamma_5 q|N\rangle&=m_Ng_5^{q,N}\langle N|\bar Ni\gamma_5 N|N\rangle,\notag\\
\langle N|\bar q\sigma^{\mu\nu} q|N\rangle&=f_{1,T}^{q,N}\langle N|\bar N\sigma^{\mu\nu} N|N\rangle,\notag\\
\langle N|\frac{\alpha_s}{4\pi}G_{\mu\nu}^a\tilde G^{\mu\nu}_a|N\rangle &= \tilde a_N\langle N|\bar N i\gamma_5 N |N\rangle,
\end{align}
we have
\begin{align}
\label{CYTL}
C_Y^{\mathcal{T},ij}&=\Big[\bar{C}^{A,i}_Y (1+\delta')^{i}\pm 2\bar{C}^{T,i}_Y\Big]\times (i\leftrightarrow j),\\
C_Y^{\mathcal{L},ij}&=\Big[\bar{C}^{A,i}_Y(1+\delta'')^i - \frac{m_\mu}{2m_N}\bar{C}^{P,i}_Y\pm 2 \bar{C}^{T,i}_Y\Big]\times (i\leftrightarrow j),\notag
\end{align}
where the upper/lower sign refers to $Y=L/R$.
For all coefficients $\bar C^N$ the isoscalar/isovector components are defined as
\beq
\bar C^0=\frac{\bar C^p+\bar C^n}{2},\qquad 
\bar C^1=\frac{\bar C^p-\bar C^n}{2},
\eeq
and $\delta'$, $\delta''$ encode the corrections from the induced pseudoscalar form factor, the axial radius, and two-body currents~\cite{Hoferichter:2020osn}---note that they are not included in the $S_{ij}^\tau$ structure factors. At $\qq^2=m_\mu^2$ they take the values $\delta'=-0.28(5)$, $\delta''=-0.44(4)$. Especially the two-body corrections lead to a sizable reduction of the $\mu\to e$ matrix elements, as also well established for nuclear $\beta$ decays~\cite{Gysbers:2019uyb}, and thus need to be included. The uncertainties are derived from the corresponding low-energy constants~\cite{Hoferichter:2015tha,Hoferichter:2015hva} and the convergence properties of the chiral expansion, as detailed in Ref.~\cite{Hoferichter:2020osn}, and also cover nuclear shell-model uncertainties, see App.~\ref{app:nuclear} for details.

The nucleon matrix elements are related by the Ward identity 
\beq
g_A^{q,N}=g_5^{q,N}-\frac{\tilde a_N}{2m_N},
\label{Ward_nucleon}
\eeq
in close analogy to Eq.~\eqref{Ward}. For the isovector combination we also keep the momentum-dependent correction from the induced pseudoscalar form factor,  which amounts to shifting $g_A^{u,p}$ and $g_A^{d,p}$ by $\mp g_A/2\times m_\mu^2/(\mpi^2+m_\mu^2)$, respectively, when applying Eq.~\eqref{Ward_nucleon} (the neutron couplings are obtained assuming isospin symmetry). Once the value of $\tilde a_N$ is determined, all $g_5^{q,N}$ thus follow from the $g_A^{q,N}$, for which we use the values from Refs.~\cite{HERMES:2006jyl,ParticleDataGroup:2020ssz,Hoferichter:2020osn} (in reasonable agreement with recent lattice-QCD calculations~\cite{Liang:2018pis,Lin:2018obj,Aoki:2021kgd}). Contrary to $a_P$, for $\tilde a_N$ only estimates based on large-$N_c$ arguments are available so far~\cite{Cheng:1988im,Cheng:2012qr}, while lattice-QCD techniques employed for the QCD $\theta$ term could allow for an ab-initio determination~\cite{Dragos:2019oxn,Bhattacharya:2021lol}. We use the estimate
\beq
\tilde a_N=-2m_N g_A^{u,0}=-0.39(12)\GeV,
\label{taN}
\eeq
with $g_A^{u,0}=(g_A^{u,p}+g_A^{u,n})/2$,
as can be derived in analogy to $a_P^\text{FKS}$ in Table~\ref{tab:matrix_elements_P}, see App.~\ref{app:GGdual}, and assign a $30\%$ uncertainty motivated by $1/N_c$ corrections.   
The tensor coefficients~\cite{Gupta:2018lvp,Hoferichter:2018zwu} are not needed as the tensor operator does not contribute to the pseudoscalar decays. 

Finally, the operators of interest for $P\to \mu e$ could also contribute to $\Br_\text{SI}[\mu\to e]$ via relativistic corrections, in analogy to the SI contribution that arises from the tensor operator at $\Order(1/m_N)$~\cite{Cirigliano:2017azj}. However, given that the matrix element of the tensor operator in  $P\to \mu e$ vanishes, such corrections are suppressed further than could be overcome by the coherent enhancement of the SI response.

\section{Limits on $\boldsymbol{P\to \mu e}$}

\begin{table}[t]
	\renewcommand{\arraystretch}{1.3}
	\centering
	\begin{tabular}{l c c c}
		\toprule
		& $\pi^0$ & $\eta$ & $\etap$\\\colrule
		$C_Y^{A,3}$ & $1.3\times 10^{-17}$
		& -- & --\\
		$C_Y^{A,8}$ & -- & $1.5\times 10^{-17}$ & $4.0\times 10^{-20}$\\
		$C_Y^{A,0}$ & -- & $2.9\times 10^{-19}$ & $2.1\times 10^{-19}$\\
		$C_Y^{P,3}$ & $4.1\times 10^{-17}$ & -- & --\\
		$C_Y^{P,8}$ & -- & $1.6\times 10^{-12}$ & $2.1\times 10^{-14}$\\
		$C_Y^{P,0}$ & -- & $4.1\times 10^{-12}$ & $5.4\times 10^{-13}$\\ 
		$C_Y^{G\tilde G}$& -- & $5.8\times 10^{-15}$ & $4.7\times 10^{-16}$\\
		\botrule
	\end{tabular}
	\caption{Limits for $\Br[P\to \mu e]\equiv\Br[P\to \mu^+ e^-+\mu^- e^+]$ that follow from $\Br[\mu\to e, \text{Ti}]<6.1\times 10^{-13}$ assuming the dominance of a single Wilson coefficient. $C_Y^{A,i}$, $C_Y^{P,i}$, $i=3,8,0$, refer to triplet, octet, and singlet components, respectively. In all cases we take $C_L=C_R$.  
	We show the worst limits obtained when scanning over the two sets of matrix elements from Table~\ref{tab:matrix_elements_P}, the couplings $g_A^{q,N}$ from Refs.~\cite{HERMES:2006jyl,Liang:2018pis,Lin:2018obj}, and $\tilde a_N$ including a $30\%$ error. The $\eta$ and $\eta'$ limits for $C_Y^{P,i}$ are sensitive to the uncertainty of $\tilde a_N$; we show the weakest limit obtained within its assigned error, but note that for the central value the limits are stronger by an order of magnitude. }    
	\label{tab:limits}
\end{table}

In general, pseudoscalar decays~\eqref{master_Pmue} and SD $\mu\to e$ conversion~\eqref{BR_SD} are not sensitive to the same linear combination of Wilson coefficients. Therefore, the translation of limits depends on the underlying BSM scenario as parameterized by the Wilson coefficients $C_Y^{A,q}$, $C_Y^{P,q}$, $C_Y^{G\tilde G}$. In the special case where only a single linear combination of Wilson coefficients contributes, the transition is immediate.
Table~\ref{tab:limits} shows the results if the triplet, octet, or singlet components of $C_Y^{A,q}$ or $C_Y^{P,q}$ are dominant, together with the case in which only $C_Y^{G\tilde G}$ is nonvanishing. The octet, singlet, and gluonic operators do not contribute to $\pi^0\to \mu e$, nor do the triplet operators to $\eta, \eta'\to\mu e$, so that considering all these flavor combinations should provide a realistic assessment of the sensitivities:
\begin{align}
\Br[\pi^0\to \mu e]&\lesssim 4\times 10^{-17},\notag\\
\Br[\eta\to \mu e]&\lesssim 4\times 10^{-12},\notag\\
\Br[\eta'\to \mu e]&\lesssim 5\times 10^{-13}.
\label{sensitivities}
\end{align}
To derive rigorous limits requires a scan over Wilson coefficients to minimize the effect in $\mu\to e$ conversion while retaining a sizable $P\to \mu e$ rate.\footnote{We set $C_L=C_R$ throughout,  as left- and right-handed components do not interfere in either rate.} Moreover, theory uncertainties due to the hadronic and nuclear matrix elements need to be taken into account. To obtain robust limits, we take the meson matrix elements either from the phenomenological or the lattice-QCD determinations quoted in Table~\ref{tab:matrix_elements_P}, similarly for the couplings $g_A^{q,N}$ from Refs.~\cite{HERMES:2006jyl,Liang:2018pis,Lin:2018obj}, and for $\tilde a_N$ as well $\delta'$, $\delta''$ we include the uncertainties as given above. All quoted limits then refer to the worst limit obtained under this variation of the hadronic and nuclear input.  

\begin{figure}[t]
	\includegraphics[width=\linewidth]{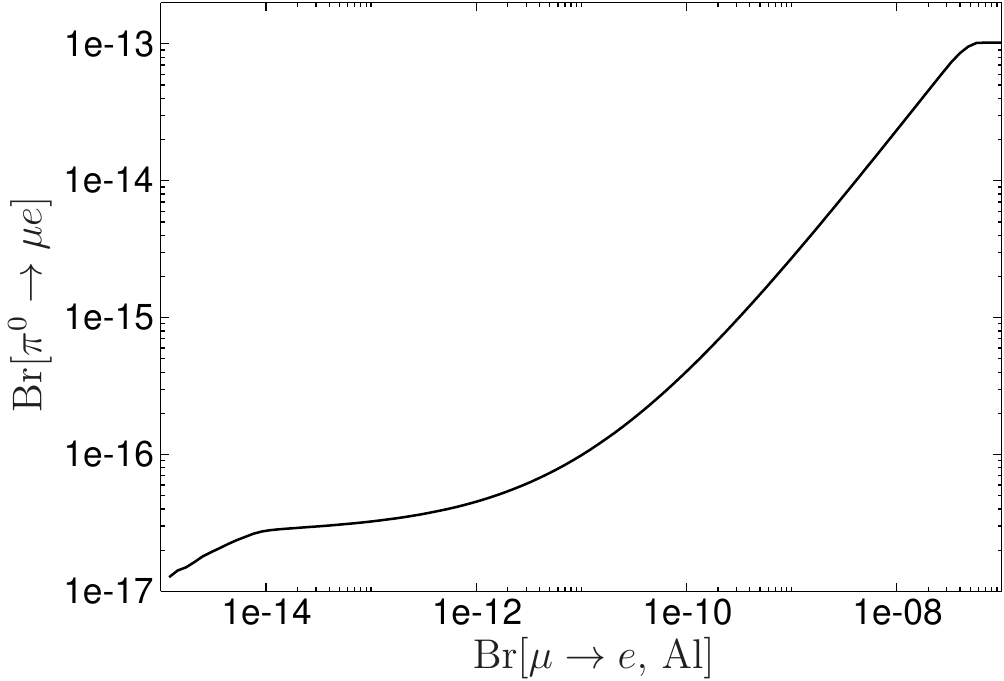}
	\caption{Limits on $\Br[\pi^0\to\mu e]$ derived from $\Br[\mu\to e, \text{Ti}]$~\cite{Wintz:1998rp} in combination with a future limit on  $\Br[\mu\to e, \text{Al}]$. Already a moderate precision for the latter suffices to substantially reduce cancellations that otherwise dilute the limit from Eq.~\eqref{sensitivities} to Eq.~\eqref{rigorous}.} 
	\label{fig:pi0}
\end{figure}

Equation~\eqref{BR_SD} shows that each multipole in the transverse and longitudinal responses is squared separately, which in $^{47}$Ti ($L=1,3,5$) and $^{49}$Ti ($L=1,3,5,7$) leads to a total of  
$2\times 3+2\times 4=14$ positive definite quantities. Accordingly, the only way to tune the rate to zero is to consider the couplings directly to protons and neutrons. Such a cancellation occurs at
\begin{align}
 C_Y^{A,u}&=C_Y^{A,d}, & C_Y^{A,s}&=-\frac{2C_Y^{A,u}g^{u,0}_A}{g_A^{s,N}},\notag\\
 \frac{C_Y^{P,u}}{m_u}&=\frac{C_Y^{P,d}}{m_d}, &\frac{C_Y^{P,s}}{m_s}&=\frac{4\pi}{\Lambda} C_Y^{G\tilde G}\frac{2g_A^{u,0}}{g_A^{u,0}-g_A^{s,N}}.
\end{align}
Since the conditions not involving strangeness remove any isovector contribution, this implies that for this choice of Wilson coefficients $\Br[\pi^0\to \mu e]$ vanishes as well. In this case, the limit is thus protected against accidental cancellation, and a scan over the parameter space establishes 
\beq
\Br[\pi^0\to \mu e]< 1.2\times 10^{-13}
\label{rigorous}
\eeq
as a rigorous limit. 
For $\eta$, $\eta'$ a nonvanishing contribution
remains, but such fine-tuned solutions are not viable due to RG corrections. As an example, we consider the dimension-$6$ contribution from $C_Y^{A,u}=C_Y^{A,d}$. If generated at a high scale $\Lambda$ above the electroweak scale $M_W$, already the  one-loop QED corrections below $M_W$ produce a vector operator~\cite{Crivellin:2017rmk,Cirigliano:2017azj}
\beq
C_Y^{V,q}\simeq -3 Q_q \frac{\alpha}{\pi}\log\frac{M_W}{m_N}C_Y^{A,q},
\eeq
with quark charges $Q_u=2/3$, $Q_d=-1/3$, and thus a contribution to the SI rate~\eqref{coefficients_SI}. 
This indirect constraint gives  
\begin{align}
\Br[\eta\to \mu e]&< 3.8\times 10^{-18},\notag\\
\Br[\etap\to \mu e]&< 9.1\times 10^{-20},
\end{align}
and thus excludes the solution via $C_Y^{A,q}$.  Therefore, a fine tuning of Wilson coefficients can relax the limits~\eqref{sensitivities}, but, realistically, only by a few orders of magnitude. Moreover, the cancellations that arise from the interference of isoscalar and isovector contributions can be substantially reduced by considering other $\mu\to e$ targets. Figure~\ref{fig:pi0} illustrates this for $\Br[\pi^0\to\mu e]$ as a function of a future limit for $\mu\to e$ conversion in Al in combination with the current Ti constraint.

\section{Conclusions}

In this Letter, we studied the connection between LFV decays of the light pseudoscalars $P=\pi^0,\eta,\eta'$ and $\mu\to e$ conversion in nuclei. The EFT approach shows that up to dimension $7$ only a few operators---axial-vector, pseudoscalar, and gluonic ones---contribute to the pseudoscalar decays, which at the same time can mediate the $\mu\to e$ conversion process albeit only by coupling to the nuclear spin. We derived master formulae for both processes to quantify their interplay, including all required hadronic matrix elements and the nuclear responses for Ti. Despite the lack of coherent enhancement for the spin-dependent response, we found that, in general, the indirect limits for $P\to \mu e$ as derived from the current $\mu\to e$ conversion limit in Ti surpass the direct ones by many orders of magnitude. Fine-tuning Wilson coefficients can relax these limits to some extent, especially for $\eta$, $\eta'$, 
but RG corrections curtail the amount of cancellations. 
The indirect limits presented here will further advance in the future with forthcoming measurements of $\mu\to e$ conversion in Al at the Mu2e and COMET experiments.

\begin{acknowledgments}
We thank Peter Wintz for valuable communication on Ref.~\cite{Wintz:1998rp}, and Vincenzo Cirigliano, Andreas Crivellin, and Bastian Kubis for helpful discussions.
This work was supported by the Swiss National Science Foundation, under Project PCEFP2\_181117, and by the ``Ram\'on y Cajal'' program with grant RYC-2017-22781, and grants CEX2019-000918-M and PID2020-118758GB-I00 funded by MCIN/AEI/10.13039/501100011033 and by ``ESF Investing in your future.''
\end{acknowledgments}

\appendix

\section{Nuclear responses}
\label{app:nuclear}

\begin{figure}[t]
	\includegraphics[width=\linewidth]{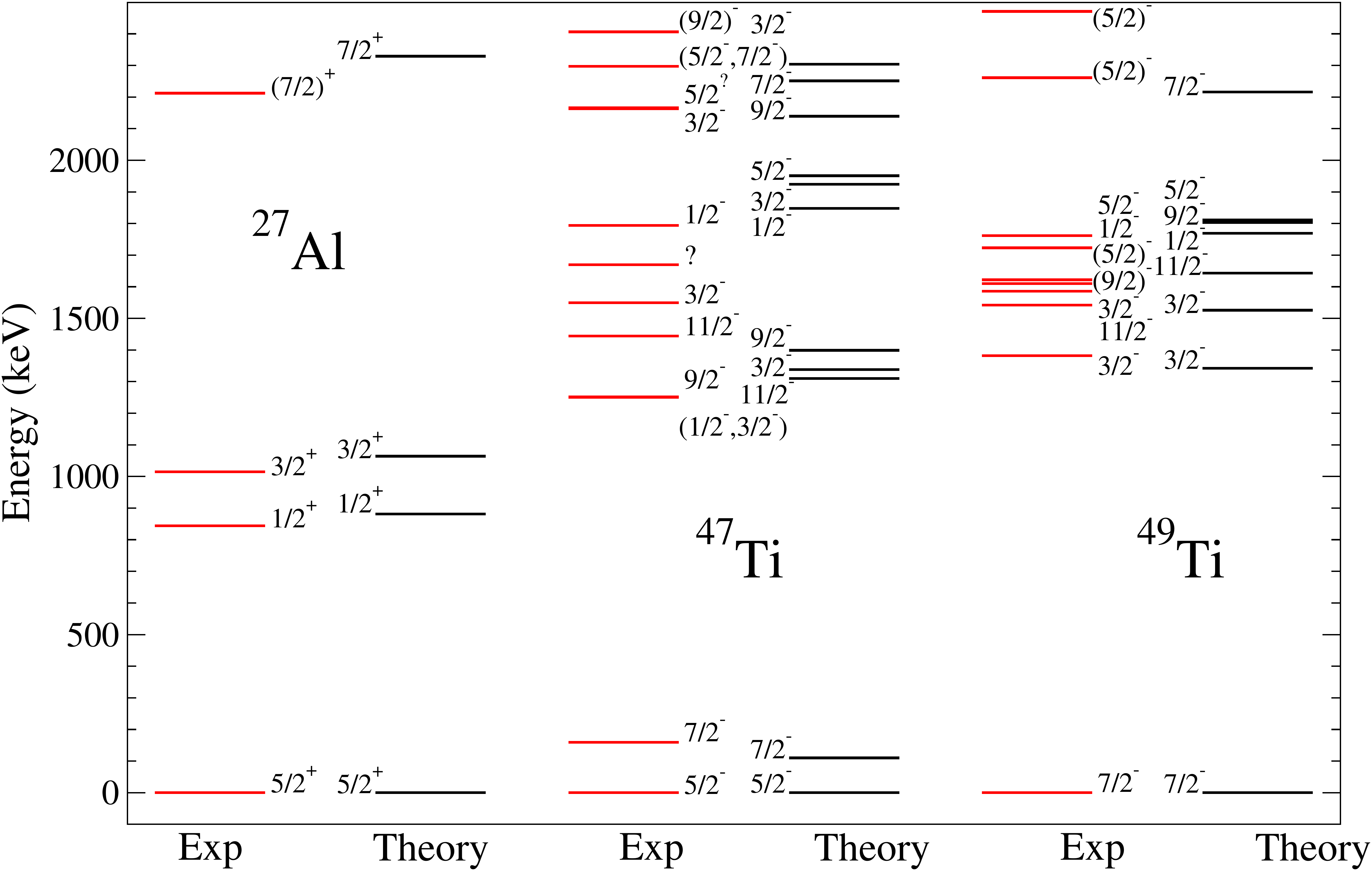}
	\caption{Calculated spectra of $^{27}$Al, $^{47}$Ti, and $^{49}$Ti compared to experimental data~\cite{ensdf}. Only positive (negative) parity states are shown for $^{27}$Al ($^{47,49}$Ti).} 
	\label{fig:spectra}
\end{figure}

Figure~\ref{fig:spectra} assesses the quality of the nuclear shell-model calculation by comparing calculated low-lying spectra of $^{47}$Ti, $^{49}$Ti, and $^{27}$Al with experiment~\cite{ensdf}. In all three cases the agreement is very good. For a comparison for $^{27}$Al including higher excitation energies, see Ref.~\cite{Klos:2013rwa}. Also, Table~\ref{tab:radii} compares calculated charge radii with experimental values and presents the shell-model results for the spin expectation values $\langle \spin_N\rangle$.

The nuclear shell-model interactions used in this work have also been used to study, in nuclei with similar mass as Al and Ti, other operators related to axial currents involving the nuclear spin such as Gamow--Teller $\beta$ decays. Calculations indicate that matrix elements are systematically overestimated, a deficiency related to the absence of two-body currents in the calculations~\cite{Gysbers:2019uyb}. Results obtained with other shell-model interactions different to the ones used in our work are in general very similar, typically within $10\%$~\cite{Richter2008,Kumar:2015zuv}. In contrast, isoscalar magnetic dipole transitions, also spin-dependent, are well reproduced in the Al region~\cite{Matsubara:2015tua}. In order to take into account these aspects, $\delta'$ and $\delta''$ in Eq.~\eqref{CYTL} include the effect of two-body currents, with an uncertainty that their effect at $|\qq|=0$, appropriate for Gamow--Teller decays, ranges from $20\%$--$30\%$. For the USDB interaction used for Al this effect is expected to be $23\%$~\cite{Richter2008} ($21\%$ for an alternative shell-model interaction in the same configuration space~\cite{Richter2008}) and for KB3G $26\%$~\cite{Martinez-Pinedo:1996zvt} ($23\%$ for an alternative shell-model interaction~\cite{Kumar:2015zuv}). Therefore, our uncertainty in $\delta'$ and $\delta''$ covers also the expected nuclear uncertainty from using alternative nuclear shell-model interactions.

Table~\ref{tab:MPhifits} summarizes the nuclear shell-model results for the $M$ and $\Phi''$ multipoles for all stable Ti and Al isotopes, while Tables~\ref{tab:Sigmafits} and~\ref{tab:SigmafitsAl} show the results for $\Sigma'$, $\Sigma''$, following the conventions from Ref.~\cite{Hoferichter:2020osn}. In all cases proton/neutron and isoscalar/isovector components are related by
$\F_\pm(\qq^2)=\F_p(\qq^2)\pm \F_n(\qq^2)$.
The SD structure factors are
\begin{align}
\label{Sij}
S_{00}^{\mathcal{T}}&= \sum_L \Big[\mathcal{F}_+^{\Sigma'_L}(\qq^2)\Big]^2, & S_{00}^{\mathcal{L}}&= \sum_L \Big[\mathcal{F}_+^{\Sigma''_L}(\qq^2)\Big]^2, \notag\\
S_{11}^{\mathcal{T}}&= \sum_L \Big[\mathcal{F}_-^{\Sigma'_L}(\qq^2)\Big]^2, & S_{11}^{\mathcal{L}}&= \sum_L \Big[\mathcal{F}_-^{\Sigma''_L}(\qq^2)\Big]^2,\notag\\
S_{01}^{\mathcal{T}}&= \sum_L 2\mathcal{F}_+^{\Sigma'_L}(\qq^2)\,\mathcal{F}_-^{\Sigma'_L}(\qq^2), & &\notag\\
S_{01}^{\mathcal{L}}&= \sum_L 2\mathcal{F}_+^{\Sigma''_L}(\qq^2)\,\mathcal{F}_-^{\Sigma''_L}(\qq^2), & &
\end{align}
with multipoles normalized as
\beq
\label{Fsigmanorm}
\mathcal{F}_N^{\Sigma'_1}(0)=\sqrt{2}\, \mathcal{F}_N^{\Sigma''_1}(0)=\sqrt{\frac{2}{3}}\sqrt{\frac{(2J+1)(J+1)}{4\pi J}} \langle \spin_N\rangle.
\eeq

\begin{table}[t!]
	\centering
	\renewcommand{\arraystretch}{1.3}
	\begin{tabular}{ccccc}
		\toprule
		& $R_\text{ch}^\text{th}$ [fm] & $R_\text{ch}^\text{exp}$ [fm] & $\langle \spin_p\rangle$ & $\langle \spin_n\rangle$\\\colrule
		
		$^{46}$Ti & $3.59$ & $3.6070(22)$ & -- &--\\
		$^{47}$Ti & $3.59$ & $3.5962(19)$ & $0.026$ & $0.293$\\
		$^{48}$Ti & $3.60$ & $3.5921(17)$ & -- &--\\
		$^{49}$Ti & $3.61$ & $3.5733(21)$ & $0.028$ & $0.367$\\
		$^{50}$Ti & $3.62$ & $3.5704(22)$ & -- &--\\
		$^{27}$Al & $3.14$ & $3.0610(31)$ & $0.326$ & $0.038$\\
		\botrule
	\end{tabular}
	\renewcommand{\arraystretch}{1.0}
	\caption{Theoretical charge radii compared to experiment from Ref.~\cite{Angeli:2013epw}, as well as calculated proton ($\langle \spin_p\rangle$) and neutron ($\langle \spin_n\rangle$) spin expectation values. 
	}
	\label{tab:radii}
\end{table}

\begin{table*}[tp]
\centering
\renewcommand{\arraystretch}{1.3}
\begin{tabular}{lcccccc}
\toprule
Isotope	& $^{46}$Ti & $^{47}$Ti & $^{48}$Ti & $^{49}$Ti  & $^{50}$Ti & $^{27}$Al\\\colrule
$J^P$ & $0^+$ & $5/2^-$ & $0^+$ & $7/2^-$ & $0^+$ & $5/2^+$\\
$\eta$ [\%] & $8.25(3)$ & $7.44(2)$ & $73.72(3)$ & $5.41(2)$ & $5.18(2)$ & $100$\\
$b$~[fm] & $1.9769$ & $1.9827$ & $1.9884$ & $1.9940$ & $1.9995$ & $1.8420$\\
\colrule
$c_1^{M+}$ & $-25.9987$ & $-27.0005$ & $-28.0021$ & $-28.9986$ & $-29.9991$ & $-11.3343$\\
$c_2^{M+}$ & $3.28239$ & $3.49161$ & $3.6798$ & $3.85975$ & $4.05453$ & $0.837814$\\
$c_3^{M+}$ & $-0.0685135$ & $-0.0797877$ & $-0.088061$ & $-0.0939764$ & $-0.102896$ & --\\\colrule
$c_1^{M-}$ & $1.99997$ & $2.99993$ & $4.00016$ & $4.99982$ & $5.99979$ & $0.666696$\\
$c_2^{M-}$ & $-0.406935$ & $-0.617487$ & $-0.830249$ & $-1.03009$ & $-1.2358$ & $-0.0858552$\\
$c_3^{M-}$ & $0.0200208$ & $0.0310325$ & $0.0424272$ & $0.0518767$ & $0.062231$ & --\\\colrule
$c_0^{\Phi''+}$ & $-4.73881$ & $-5.91242$ & $-6.77176$ & $-7.96954$ & $-8.91559$ & $-4.98975$\\
$c_1^{\Phi''+}$ & $1.8957$ & $2.365$ & $2.70879$ & $3.18778$ & $3.566129$ & $0.997985$\\
$c_2^{\Phi''+}$ & $-0.140202$ & $-0.175856$ & $-0.198666$ & $-0.231598$ & $-0.25760$ & --\\\colrule
$c_0^{\Phi''-}$ & $1.61434$ & $2.67922$ & $3.45274$ & $4.5949$ & $5.54649$ & $0.303398$\\
$c_1^{\Phi''-}$ & $-0.645785$ & $-1.07173$ & $-1.38116$ & $-1.8379$ & $-2.2185$ & $-0.0606814$\\
$c_2^{\Phi''-}$ & $0.0461341$ & $0.0771383$ & $0.100134$ & $0.133188$ & $0.160647$ & --\\\botrule
\end{tabular}
\renewcommand{\arraystretch}{1.0}
\caption{Spin/parity $J^P$ of the nuclear ground states, natural abundance $\eta$, 
harmonic-oscillator length $b$, and fit coefficients for the nuclear multipoles $\F_\pm^M$ and $\F_\pm^{\Phi''}$.  The parameterizations $\F_\pm^{M}(u) = e^{-\frac{u}{2}}\sum_{i=0}^{3} c_i^{M\pm} u^i$
(with $c_0^{M+}=Z+N$ and $c_0^{M-}=Z-N$) and $\F_{\pm}^{\Phi''}(u) =
e^{-\frac{u}{2}} \sum_{i=0}^{2} c_i^{\Phi''\pm} u^i$, with $u=\qq^2b^2/2$, correspond to the analytical form in the harmonic-oscillator
basis~\cite{Donnelly:1979ezn}.}
\label{tab:MPhifits}
\end{table*}

\begin{table*}[tp]
\centering
\renewcommand{\arraystretch}{1.3}
\begin{tabular}{lccccccc}
\toprule
Isotope	& \multicolumn{3}{c}{$^{47}$Ti} & \multicolumn{4}{c}{$^{49}$Ti}\\
$L$ & $1$ & $3$ &$5$ & $1$ & $3$ &$5$ & $7$\\
\colrule
$c_0^{\Sigma'p}$ & $0.0175579$ & -- & -- & $0.0203333$ & -- & --& -- \\
$c_1^{\Sigma'p}$ & $-0.0179636$ & $0.00453093$ & -- & $-0.0214563$ & $-0.00759501$ & --& -- \\
$c_2^{\Sigma'p}$ & $0.00314365$ & $-0.00134858$ & $-0.0000163753$ & $0.00397767$ & $0.00354166$ & $-0.00043197$& -- \\
$c_3^{\Sigma'p}$ & $-0.0000982683$ & $0.0000307488$ & $0.0000216414$ & $-0.000155929$ & $-0.000272126$ & $0.0000408714$& $-2.75178\times 10^{-6}$ \\\colrule
$c_0^{\Sigma'n}$ & $0.19529$ & -- & -- & $0.271019$ & -- & --& -- \\
$c_1^{\Sigma'n}$ & $-0.245141$ & $0.0279757$ & -- & $-0.357078$ & $-0.129809$ & --& -- \\
$c_2^{\Sigma'n}$ & $0.0551808$ & $-0.0120502$ & $-0.0298524$ & $0.0813202$ & $0.0399765$ & $0.0179357$& -- \\
$c_3^{\Sigma'n}$ & $-0.00288588$ & $0.0010701$ & $0.00255996$ & $-0.00432571$ & $-0.00229466$ & $-0.00108624$& $-0.00576536$ \\\colrule
$c_0^{\Sigma''p}$ & $0.0124152$ & -- & -- & $0.0143777$ & -- & --& -- \\
$c_1^{\Sigma''p}$ & $-0.0118426$ & $0.00392391$ & -- & $-0.0127874$ & $-0.00657747$ & --& -- \\
$c_2^{\Sigma''p}$ & $0.00272966$ & $-0.000701449$ & $-0.0000149486$ & $0.00252953 $ & $0.00196317$ & $-0.000394335$& -- \\
$c_3^{\Sigma''p}$ & $-0.000176902$ & $0.0000103558$ & $0.0000122962$ & $-0.000122398$ & $-0.000120552$ & $0.0000438309$& $-2.57402\times 10^{-6}$ \\\colrule
$c_0^{\Sigma''n}$ & $0.138089$ & -- & -- & $0.191638$ & -- & --& -- \\
$c_1^{\Sigma''n}$ & $-0.0676043$ & $0.0242274$ & -- & $-0.0699121$ & $-0.112418$ & --& -- \\
$c_2^{\Sigma''n}$ & $0.00783041$ & $-0.00458878$ & $-0.0272513$ & $0.00435709$ & $0.0179024$ & $0.0163729$& -- \\
$c_3^{\Sigma''n}$ & $-0.000298162$ & $0.000464503$ & $0.00168688$ & $0.0000212473$ & $-0.000600979$ & $-0.000188205$& $-0.00539299$ \\\botrule
\end{tabular}
\renewcommand{\arraystretch}{1.0}
\caption{Fit coefficients for the nuclear
multipoles $\F_{N}^{\Sigma'_L}$ and $\F_{N}^{\Sigma''_L}$ of $^{47,49}$Ti, using the fit
function $\F(u) = e^{-\frac{u}{2}}\sum_{i=0}^3 c_i u^i$.}
\label{tab:Sigmafits}
\end{table*}

\begin{table}[t]
\centering
\renewcommand{\arraystretch}{1.3}
\begin{tabular}{lccc}
\toprule
Isotope	& \multicolumn{3}{c}{$^{27}$Al}\\
$L$ & $1$ & $3$ &$5$\\
\colrule
$c_0^{\Sigma'p}$ & $0.217688$ & -- & --\\
$c_1^{\Sigma'p}$ & $-0.199127$ & $-0.0543796$ & --\\
$c_2^{\Sigma'p}$ & $0.0274599$ & $0.00651646$ & $0.0277341$\\
\colrule
$c_0^{\Sigma'n}$ & $0.0255784$ & -- & --\\
$c_1^{\Sigma'n}$ & $-0.00985665$ & $0.000280434$& -- \\
$c_2^{\Sigma'n}$ & $-0.00150251$ & $-0.000488419$ & $0.000128524$ \\
\colrule
$c_0^{\Sigma''p}$ & $0.153928$ & -- & -- \\
$c_1^{\Sigma''p}$ & $-0.0262479$ & $-0.0470943$ &  -- \\
$c_2^{\Sigma''p}$ & $0.00103769$ & $-0.00206298$ & $0.0253177$\\\colrule
$c_0^{\Sigma''n}$ & $0.0180868$ & -- & -- \\
$c_1^{\Sigma''n}$ & $-0.022234$ & $0.000242867$ & -- \\
$c_2^{\Sigma''n}$ & $0.00278141$ & $0.000103865$ & $0.000117327$ \\\botrule
\end{tabular}
\renewcommand{\arraystretch}{1.0}
\caption{Same as Table~\ref{tab:Sigmafits} for $^{27}$Al.}
\label{tab:SigmafitsAl}
\end{table}

\section{Matrix elements of $\boldsymbol{G\tilde G}$}
\label{app:GGdual}

At present, no model-independent determination of the nucleon matrix element of the $G^a_{\mu\nu} \tilde{G}_a^{\mu\nu}$ operator is available, and an additional condition is necessary to solve for all couplings $g_5^{q,N}$, $\tilde a_N$ in Eq.~\eqref{Ward_nucleon}. As suggested by large-$N_c$ arguments, frequently one imposes~\cite{Cheng:2012qr}
\beq
\label{N_gamma_5}
\langle N|\sum_q\bar q i\gamma_5 q|N\rangle=0.
\eeq

As a first step, we study the consequences of the analog condition for the case of the pseudoscalar matrix elements. Imposing  
 $\sum_q b_q h_P^q/m_q=0$, we solve Eq.~\eqref{Ward} to obtain
\beq
a_P^\text{FKS}=-\frac{M_P^2(b_uf_P^u+b_d f_P^d)}{2},
\label{a_P_FKS}
\eeq
where we expanded in $m_{u,d}\ll m_s$ and assumed the isospin limit, as throughout this work. The approximation~\eqref{a_P_FKS} coincides with the result in the FKS $\eta$--$\eta'$
mixing scheme: starting from~\cite{Beneke:2002jn}  
\begin{align}
 a_\eta^\text{FKS}&=-\frac{M_\etap^2-M_\eta^2}{2\sqrt{2}}\sin(2\phi)\Big[-f_q \sin\phi+\sqrt{2} f_s\cos\phi\Big],\notag\\
 a_\etap^\text{FKS}&=-\frac{M_\etap^2-M_\eta^2}{2\sqrt{2}}\sin(2\phi)\Big[f_q \cos\phi+\sqrt{2} f_s\sin\phi\Big],
\end{align}
with 
\begin{align}
f_\eta^u&=f_\eta^d=f_q\cos \phi, &
f_\eta^s&=-f_s\sin\phi,\notag\\
f_\etap^u&=f_\etap^d=f_q\sin \phi, &
f_\etap^s&=f_s\cos\phi,
\end{align}
expressed in terms of a single mixing angle $\phi$, the expression~\eqref{a_P_FKS} is indeed reproduced upon
using~\cite{Feldmann:1998vh}
\beq
\frac{\sqrt{2}f_s}{f_q}=\tan(\phi-\theta_8),\qquad \cot\theta_8=-\frac{M_\etap^2}{M_\eta^2}\tan\phi,
\eeq
with the latter also assuming the $m_{u,d}\ll m_s$ limit. Table~\ref{tab:matrix_elements_P} indicates that the results obtained in this way agree well with the lattice calculation of Ref.~\cite{Bali:2021qem}. Since the corrections to the FKS scheme are indeed expected to be suppressed in the large-$N_c$ limit, it appears natural that the large-$N_c$ arguments from  Ref.~\cite{Cheng:2012qr} lead to the same result. In addition, the numerical agreement for the pseudoscalar matrix elements suggests that the corresponding estimate for the nucleon case~\eqref{taN} is reasonable as well. Note that, in contrast to Ref.~\cite{Cheng:2012qr}, we evaluate this estimate again in the isospin limit, since corrections can only be assessed in a consistent manner considering also isospin-breaking effects in the axial-vector couplings $g_A^{q,N}$.

\bibliography{Pmue}

\end{document}